\documentclass[preprintnumbers,amsmath,amssymb,floatfix,11pt,
superscriptaddress,nofootinbib]{revtex4}
\usepackage{color}

\begin{document}

\title{Correspondence between entropy-corrected holographic and
Gauss-Bonnet dark energy models}

\author{ M. R. Setare} \email{rezakord@ipm.ir}
\affiliation{Department of Science, Payame Noor University, Bijar,
Iran} \affiliation{Research Institute for Astronomy and Astrophysics
of Maragha (RIAAM), Maragha, Iran }

\author{ Mubasher Jamil} \email{mjamil@camp.nust.edu.pk}
\affiliation{Center for Advanced Mathematics and Physics, National
University of Sciences and Technology, H-12, Islamabad, Pakistan}

\begin{abstract}
\vspace*{1.5cm} \centerline{\bf Abstract} \vspace*{1cm} In the
present work we investigate the cosmological implications of the
entropy-corrected holographic dark energy (ECHDE) density in the
Gauss-Bonnet framework. This is motivated from the loop quantum
gravity corrections to the entropy-area law. Assuming the two
cosmological scenarios are valid simultaneously, we show that there
is a correspondence between the ECHDE scenario in flat universe and
the phantom dark energy model in the framework of Gauss-Bonnet
theory with a potential. This correspondence leads consistently to
an accelerating universe.
\end{abstract}
\maketitle
\newpage

\section{Introduction}
Recent cosmological observations indicate that our universe is in
accelerated expansion. These observations are those which is
obtained by SNe Ia {\cite{c1}}, WMAP {\cite{c2}}, SDSS {\cite{c3}}
and X-ray {\cite{c4}}.  These observations also suggest that our
universe is spatially flat, and consists of about $70 \%$ dark
energy (DE) with negative pressure, $30\%$ dust matter (cold dark
matter plus baryons), and negligible radiation. In order to explain
why the cosmic acceleration happens, many theories have been
proposed. The simplest candidate of the dark energy is a tiny
positive time-independent cosmological constant $\Lambda$, for which
$\omega=-1$. However, it is difficult to understand why the
cosmological constant is about 120 orders of magnitude smaller than
its natural expectation (the Planck energy density). This is the
so-called cosmological constant problem. Another puzzle of the dark
energy is the cosmological coincidence problem: why are we living in
an epoch in which the dark energy density and the dust matter energy
are comparable?. An alternative proposal for dark energy is the
dynamical dark energy scenario.
 The dynamical nature of dark
energy, at least in an effective level, can originate from various
fields, such is a canonical scalar field (quintessence)
\cite{quint}, a phantom field, that is a scalar field with a
negative sign of the kinetic term \cite{phant}, or the combination
of quintessence and phantom in a unified model named quintom
\cite{quintom}. Recently another paradigm has been
constructed in the light of the holographic principle of quantum
gravity theory, and thus it presents some interesting features of
an underlying theory of dark energy \cite{holoprin}. This paradigm may
simultaneously provide a solution to the coincidence
problem \cite{Li:2004rb}. The holographic dark energy model has been
extended to include the spatial curvature contribution
\cite{nonflat} and it has also been generalized in the braneworld
framework \cite{bulkhol}. Lastly, it has been tested and
constrained by various astronomical observations
\cite{obs3,HG,observHDExray,observHDECMB,observHDE}.
Since holographic energy density corresponds to a dynamical
cosmological constant, we need a dynamical framework, instead of
general relativity, to consistently accommodate it. A proposal, closely
related to the low-energy
string effective action, is the scalar-Gauss-Bonnet gravity
\cite{3}, which can be considered as a form of gravitational dark
energy.\\
In the present paper we are interested in investigating the
conditions under which we can obtain a correspondence between
holographic and Gauss-Bonnet models of dark energy, i.e to examine
holographic dark energy in a spatially flat Gauss-Bonnet universe.

\section{Gauss-Bonnet Dark Energy}

In this section we provide the basic Gauss-Bonnet model for dark
energy \cite{3,4,Nojiri2}. In this framework, the candidate for dark
energy is a scalar field $\phi$, which is moreover coupled to
gravity through the higher-derivative (string-originated)
Gauss-Bonnet term. The corresponding action is given by
\begin{equation}
S=\int d^{4}x\,\sqrt{g}\,\left[
\frac{1}{2\kappa^2}R-\frac{\sigma}{2}\partial_{\mu}\phi\partial^{\mu}\phi-V(\phi)+f(\phi)G'\right]
, \label{action}
\end{equation}
where $\kappa^2=8\pi G$ and $\sigma=\pm1$. Also $f(\phi)$ is an
arbitrary function of $\phi$ which denotes the coupling of the field
with the geometry. For the sake of generality, we consider both
behaviors of the scalar field i.e. canonical scalar field
$\sigma=1$, and $\sigma=-1$ which corresponds to phantom behavior.
In the above expression (\ref{action}), the quantity $G'$ represents
the Gauss-Bonnet term which is explicitly written as:
\begin{equation}
G'\equiv
  R^{2}-4R_{\mu\nu}R^{\mu\nu}+R_{\mu\nu\rho\sigma}R^{\mu\nu\rho\sigma},
\label{GB}
\end{equation}
where $R_{\mu\nu\rho\sigma}$, $R_{\mu\nu}$ and $R$ are respectively
the Riemann and Ricci tensors and $R$ is the curvature scalar while
$g_{\mu\nu}$ is the background metric. Motivated by several
observational and empirical findings \cite{c1,c2,c3,c4}, we shall
focus on the spatially flat Robertson-Walker universe with
\begin{equation}\label{met}
ds^{2}=-dt^{2}+a(t)^{2}(dr^{2}+r^{2}d\Omega^{2}),
\end{equation}
where we took $k=0$ in (\ref{action}).

The equation of motion for the scale factor is \cite{4}:
\begin{equation}\label{2}
\frac{\sigma}{2}\dot{\phi}^{2}-V(\phi)+16f'(\phi)\dot{\phi}H\frac{\ddot{a}}{a}
+8\left[f'(\phi)\ddot{\phi}
+f''(\phi)\dot{\phi}^{2}\right]H^2=p_\Lambda,
\end{equation}
while for the scalar field, we have
\begin{equation}\label{3}
\sigma\left[\ddot{\phi}+3H\dot{\phi}+\frac{V'(\phi)}{\sigma}\right]=24f'(\phi)H^2\frac{\ddot{a}}{a}.
\end{equation}
Moreover we have a constraint equation, namely:
\begin{equation}\label{1}
\frac{\sigma}{2}\dot{\phi}^{2}+V(\phi)-24f'(\phi)\dot{\phi}H^3=\rho_\Lambda.
\end{equation}
In the expressions (\ref{2}) and (\ref{1}) above, $p_\Lambda$ and
$\rho_\Lambda$ are the pressure and energy density due to the scalar
field and the Gauss Bonnet interaction \cite{Nojiri2}, which are
identified as the corresponding quantities of dark energy.

\section{Entropy corrected holographic dark energy}

The black hole entropy plays a central role in the derivation of
holographic dark energy (HDE). Indeed, the definition and derivation
of holographic energy density depends on the entropy-area
relationship $S\sim A \sim L^2$ of black holes in Einstein's
gravity, where $A \sim L^2$ represents the area of the horizon.
However, this definition can be modified from the inclusion of
quantum effects, motivated from the loop quantum gravity (LQG). The
quantum corrections provided to the entropy-area relationship leads
to the curvature correction in the Einstein-Hilbert action and vice
versa \cite{13}. The corrected entropy takes the form \cite{14}
\begin{equation}\label{entropy}
S=\frac{A}{4}+\tilde\gamma \ln\Big(\frac{A}{4}\Big)+\tilde\beta,
\end{equation}
where $\tilde\gamma$ and $\tilde\beta$ are dimensionless constants
of order unity. The exact values of these constants are not yet
determined and still debatable in loop quantum cosmology. These
corrections arise in the black hole entropy in LQG due to thermal
equilibrium fluctuations and quantum fluctuations \cite{15}. It is
very interesting if one can determine the coefficient in front of
log correction term by observational constraints. This term also
appears in a model of entropic cosmology which unifies the inflation
and late time acceleration, see \cite{cai}, and it was found the
coefficient might be extremely large due to current cosmological
constraint, which inevitably brought a fine tuning problem to
entropy corrected models. Taking the corrected entropy-area relation
(\ref{entropy}) into account, the energy density of the HDE will be
modified as well. On this basis, Wei \cite{16} proposed the energy
density of the so-called ``entropy-corrected holographic dark
energy'' (ECHDE) in the form
\begin{equation}
 \rho_{\Lambda}=3c^2R_h^{-2}+\gamma R_h^{-4}\ln(R_h^{2})+\beta
 R_h^{-4},
\label{holo}
\end{equation}
in units where $M_p^2=8\pi G=1$, and $c$
is a constant which value is determined by observational fit. The
future event horizon $R_h$ is defined as
\begin{equation}
R_h= a\int_t^\infty \frac{dt}{a}=a\int_a^\infty\frac{da}{Ha^2},
\label{Rh}
\end{equation}
which leads to results compatible with observations. Furthermore, we
can define the dimensionless dark energy as:
\begin{equation}
\Omega_{\Lambda}\equiv\frac{\rho_{\Lambda}}{3H^2}=\frac{3c^2+\gamma
R_h^{-2}\ln(R_h^2)+\beta R_h^{-2}}{3H^2R_h^2}\label{omega}.
\end{equation}
In the case of a dark-energy dominated universe, dark energy
evolves according to the conservation law
\begin{equation}
\dot{\rho}_{\Lambda}+3H(\rho_{\Lambda}+p_{\Lambda})=0
\label{coneq},
\end{equation}
or equivalently
\begin{equation}
\dot{\Omega}_\Lambda=-\frac{2\dot H}{3H^3R_h^2}(3c^2+\gamma
R_h^{-2}\ln(R_h^{2}+\beta R_h^{-2})+\frac{HR_h-1}{3H^2R_h^3}\Big[
-6c^2+2\gamma R_h^{-2}-4\gamma R_h^{-2}\ln R_h^{2}-4\beta R_h^{-3}
\Big],
\end{equation}
where the equation of state is
\begin{equation}
p_{\Lambda}=\left[-1-\frac{ 2\gamma R_h^{-2} -4\gamma R_h^{-2}\ln
(R_h^{2})-4\beta R_h^{-2}-6c^2}{3(3c^2+\gamma R_h^{-2}\ln
(R_h^{2})+\beta
R_h^{-2})}\left\{1-\sqrt{\frac{3\Omega_\Lambda}{3c^2+\gamma
R_h^{-2}\ln(R_h^2)+\beta R_h^{-2}}}\right\}\right]\rho_{\Lambda}
\label{eqstat},
\end{equation}
 which leads straightforwardly to an
index of the equation of state of the form:
\begin{equation}\label{index}
w_{\Lambda }=-1-\frac{ 2\gamma R_h^{-2} -4\gamma R_h^{-2}\ln
(R_h^{2})-4\beta R_h^{-2}-6c^2}{3(3c^2+\gamma R_h^{-2}\ln
(R_h^{2})+\beta
R_h^{-2})}\left[1-\sqrt{\frac{3\Omega_\Lambda}{3c^2+\gamma
R_h^{-2}\ln(R_h^2)+\beta R_h^{-2}}}\right].
\end{equation}

\section{Correspondence between ECHDE and
Gauss-Bonnet Dark Energy models}

We want to obtain the conditions under which there is a
correspondence between the Gauss-Bonnet dark energy model and the
entropy corrected holographic dark energy scenario, in the flat
background space. In particular, to determine an appropriate
Gauss-Bonnet potential which makes the two pictures to coincide with
each other.

Let us first consider the simple Gauss-Bonnet solutions acquired
in \cite{4,Nojiri2}. In this case $f(\phi)$ is given as \cite{3}
\begin{equation}
\label{fphi} f(\phi)=f_0e^{\frac{2\phi}{\phi_{0}}}.
\end{equation}
In addition, we assume that the scale factor behaves as
$a=a_0t^{h_0}$, and similarly to \cite{4} we will examine both
$h_0$-sign cases. However, when $h_0$ is negative the scale factor
does not correspond to expanding universe but to shrinking one. If
one changes the direction of time as $t\rightarrow -t$, the
expanding universe whose scale factor is given by $a=a_0(-t)^{h_0}$
emerges \footnote{ For this form of scale factor one could obtain an
interesting phenomenon when $t$ arrives $t_s$, i.e., a big rip
singularity \cite{jims}.  So this is an important scenario and also
its relation with other cosmological singularities \cite{cai2}.}.
Since $h_0$ is not an integer in general, there is one remaining
difficulty concerning the sign of $t$. To avoid the apparent
inconsistency, we may further shift the origin of the time as
$t\rightarrow -t\rightarrow t_s-t$. Then the time $t$ can be
positive as long as $t < t_s$, and we can consistently write
$a=a_0(t_s-t)^{h_0}$. Thus, we can finally write \cite{4}
\begin{equation}
\label{h0}
 H=\frac{h_0}{t},
\hspace{1cm}\phi=\phi_{0} \ln \frac{t}{t_1}
 \end{equation}
 when
$h_0> 0$ or
\begin{equation}
\label{h01} H=\frac{-h_0}{t_s-t}, \hspace{1cm}\phi=\phi_{0}
\ln \frac{t_s-t}{t_1}
\end{equation}
when $h_0< 0$, with an
undetermined constant $t_1$.

Let us first investigate the positive-$h_0$ case. If we establish
a correspondence between the holographic dark energy and
Gauss-Bonnet approach, then using dark energy density equation
(\ref{1}) and relation (\ref{omega}), together with expressions
(\ref{h0}), we can easily derive the scalar potential term as
\begin{equation}
\label{Vphi}
V=\frac{e^{-\frac{2\phi}{\phi_{0}}}}{t_1^{2}}\left(3\Omega_{\Lambda}h_{0}^{2}
+\frac{48f_0h_{0}^{3}}{t_1^{2}} -\frac{\sigma \phi_0^{2}}{2}\right).
 \end{equation}
Note that expressions (\ref{h0}) allow for an elimination of time
$t$ in terms of the scalar field $\phi$. Furthermore, by
substituting $\phi$, and $H$ from (\ref{h0}), $f(\phi)$ from
(\ref{fphi}) and $V(\phi)$ from (\ref{Vphi}) into (\ref{3}) we
obtain:
\begin{equation}
\label{7}
 -3\sigma
h_0\phi_{0}+\frac{6\Omega_{\Lambda}h_0^{2}}{\phi_{0}}+\frac{96f_0h_0^{3}}{\phi_{0}
t_1^{2}}-3h_0^{2}\frac{d\Omega_{\Lambda}}{d\phi}+\frac{48f_0h_0^{3}(h_0-1)}{\phi_{0}
t_1^{2}}=0
\end{equation}
where
\begin{equation}
\frac{d\Omega_{\Lambda}}{d\phi}=\frac{d\Omega_{\Lambda}}{dt}\frac{t}{\phi_{0}}=\frac{d\Omega_{\Lambda}}
{dt}\frac{t_1}{\phi_{0}}\,e^{\frac{\phi}{\phi_{0}}}. \label{8}
\end{equation}
Now, under the ansatz $a=a_0t^{h_0}$ it is easy to see from
(\ref{Rh}) that in order for $R_h$ to be finite, $h_0$ has to be
greater than 1. In such a case we straightforwardly find:
\begin{equation}
  R_h=\frac{t}{h_0-1},
\end{equation}
\begin{equation}
\Omega_\Lambda=\frac{(h_0-1)^2}{3h_0^2}\Big[3c^2+\gamma
\Big(\frac{t_1e^{\frac{\phi}{\phi_{0}}}}{h_0-1}\Big)^{-2}\ln\{\Big(\frac{t_1e^{\frac{\phi}{\phi_{0}}}}{h_0-1}\Big)^2\}+\beta
\Big(\frac{t_1e^{\frac{\phi}{\phi_{0}}}}{h_0-1}\Big)^{-2}\Big],
\end{equation}
and
\begin{eqnarray}
w_{\Lambda }&=&-1-\frac{ 2\gamma
\Big(\frac{t_1e^{\frac{\phi}{\phi_{0}}}}{h_0-1}\Big)^{-2} -4\gamma
\Big(\frac{t_1e^{\frac{\phi}{\phi_{0}}}}{h_0-1}\Big)^{-2}\ln
\Big\{\Big(\frac{t_1e^{\frac{\phi}{\phi_{0}}}}{h_0-1}\Big)^{2}\Big\}-4\beta
\Big(\frac{t_1e^{\frac{\phi}{\phi_{0}}}}{h_0-1}\Big)^{-2}-6c^2}{3(3c^2+\gamma
\Big(\frac{t_1e^{\frac{\phi}{\phi_{0}}}}{h_0-1}\Big)^{-2}\ln
\Big\{\Big(\frac{t_1e^{\frac{\phi}{\phi_{0}}}}{h_0-1}\Big)^{2}\Big\}+\beta
\Big(\frac{t_1e^{\frac{\phi}{\phi_{0}}}}{h_0-1}\Big)^{-2})}\nonumber\\&&\times\left[1-\sqrt{\frac{3\Omega_\Lambda}{3c^2+\gamma
\Big(\frac{t_1e^{\frac{\phi}{\phi_{0}}}}{h_0-1}\Big)^{-2}\ln\Big\{\Big(\frac{t_1e^{\frac{\phi}{\phi_{0}}}}{h_0-1}\Big)^2\Big\}+\beta
\Big(\frac{t_1e^{\frac{\phi}{\phi_{0}}}}{h_0-1}\Big)^{-2}}}\right].\label{wh0}
\end{eqnarray}

Let us proceed to the investigation of the negative-$h_0$ case.
Repeating the same steps, but imposing relations (\ref{h01}) we
find that
\begin{equation}
\label{Vphi1}
V=\frac{e^{-\frac{2\phi}{\phi_{0}}}}{t_1^{2}}\left(3\Omega_{\Lambda}h_{0}^{2}
+\frac{48f_0h_{0}^{3}}{t_1^{2}} -\frac{\sigma \phi_0^{2}}{2}\right),
 \end{equation}
and
\begin{equation}
\label{conb} -2\sigma \phi_{0} -3\sigma
h_0\phi_{0}+\frac{6\Omega_{\Lambda}h_0^{2}}{\phi_{0}}-\frac{96f_0h_0^{3}}{\phi_{0}
t_1^{2}}-3h_0^{2}\frac{d\Omega_{\Lambda}}{d\phi}+\frac{48f_0h_0^{3}(h_0-1)}{\phi_{0}
t_1^{2}}=0,
\end{equation}
where
\begin{equation}\frac{d\Omega_{\Lambda}}{d\phi}=-\frac{d\Omega_{\Lambda}}{dt}\frac{(t_s-t)}{\phi_{0}}=-\frac{d\Omega_{\Lambda}}
{dt}\frac{t_1}{\phi_{0}}\,e^{\frac{\phi}{\phi_{0}}}. \label{81}
\end{equation}
Now, under the ansatz $a=a_0(t_s-t)^{h_0}$ we can see from
(\ref{Rh}) that $R_h$ is always finite if $h_0$ is negative, which
is just the case under investigation. Then we have:
\begin{equation}
R_h=\frac{t_s-t}{1-h_0},
\end{equation}
\begin{equation}
\Omega_\Lambda=\frac{(h_0-1)^2}{3h_0^2}\Big[3c^2+\gamma
\Big(\frac{t_1e^{\frac{\phi}{\phi_{0}}}}{h_0-1}\Big)^{-2}\ln\{\Big(\frac{t_1e^{\frac{\phi}{\phi_{0}}}}{h_0-1}\Big)^2\}+\beta
\Big(\frac{t_1e^{\frac{\phi}{\phi_{0}}}}{h_0-1}\Big)^{-2}\Big],
\end{equation}
and therefore
\begin{eqnarray}
w_{\Lambda }&=&-1-\frac{ 2\gamma
\Big(\frac{t_1e^{\frac{\phi}{\phi_{0}}}}{h_0-1}\Big)^{-2} -4\gamma
\Big(\frac{t_1e^{\frac{\phi}{\phi_{0}}}}{h_0-1}\Big)^{-2}\ln
\Big\{\Big(\frac{t_1e^{\frac{\phi}{\phi_{0}}}}{h_0-1}\Big)^{2}\Big\}-4\beta
\Big(\frac{t_1e^{\frac{\phi}{\phi_{0}}}}{h_0-1}\Big)^{-2}-6c^2}{3(3c^2+\gamma
\Big(\frac{t_1e^{\frac{\phi}{\phi_{0}}}}{h_0-1}\Big)^{-2}\ln
\Big\{\Big(\frac{t_1e^{\frac{\phi}{\phi_{0}}}}{h_0-1}\Big)^{2}\Big\}+\beta
\Big(\frac{t_1e^{\frac{\phi}{\phi_{0}}}}{h_0-1}\Big)^{-2})}\nonumber\\&&\times\left[1-\sqrt{\frac{3\Omega_\Lambda}{3c^2+\gamma
\Big(\frac{t_1e^{\frac{\phi}{\phi_{0}}}}{h_0-1}\Big)^{-2}\ln\Big\{\Big(\frac{t_1e^{\frac{\phi}{\phi_{0}}}}{h_0-1}\Big)^2\Big\}+\beta
\Big(\frac{t_1e^{\frac{\phi}{\phi_{0}}}}{h_0-1}\Big)^{-2}}}\right].\label{wh01}
\end{eqnarray}
We also mention the conditions under which phantom crossing
\cite{rev} can be realized in the present scenario:
\begin{equation}\label{c1}
2\gamma \Big(\frac{t_1e^{\frac{\phi}{\phi_{0}}}}{h_0-1}\Big)^{-2}
-4\gamma
\Big(\frac{t_1e^{\frac{\phi}{\phi_{0}}}}{h_0-1}\Big)^{-2}\ln
\Big\{\Big(\frac{t_1e^{\frac{\phi}{\phi_{0}}}}{h_0-1}\Big)^{2}\Big\}-4\beta
\Big(\frac{t_1e^{\frac{\phi}{\phi_{0}}}}{h_0-1}\Big)^{-2}-6c^2>0,
\end{equation}
\begin{equation}\label{c2}
3\Omega_\Lambda<3c^2+\gamma
\Big(\frac{t_1e^{\frac{\phi}{\phi_{0}}}}{h_0-1}\Big)^{-2}\ln\Big\{\Big(\frac{t_1e^{\frac{\phi}{\phi_{0}}}}{h_0-1}\Big)^2\Big\}+\beta
\Big(\frac{t_1e^{\frac{\phi}{\phi_{0}}}}{h_0-1}\Big)^{-2}.
\end{equation}
and
\begin{equation}\label{c3}
2\gamma \Big(\frac{t_1e^{\frac{\phi}{\phi_{0}}}}{h_0-1}\Big)^{-2}
-4\gamma
\Big(\frac{t_1e^{\frac{\phi}{\phi_{0}}}}{h_0-1}\Big)^{-2}\ln
\Big\{\Big(\frac{t_1e^{\frac{\phi}{\phi_{0}}}}{h_0-1}\Big)^{2}\Big\}-4\beta
\Big(\frac{t_1e^{\frac{\phi}{\phi_{0}}}}{h_0-1}\Big)^{-2}-6c^2<0,
\end{equation}
\begin{equation}\label{c4}
3\Omega_\Lambda>3c^2+\gamma
\Big(\frac{t_1e^{\frac{\phi}{\phi_{0}}}}{h_0-1}\Big)^{-2}\ln\Big\{\Big(\frac{t_1e^{\frac{\phi}{\phi_{0}}}}{h_0-1}\Big)^2\Big\}+\beta
\Big(\frac{t_1e^{\frac{\phi}{\phi_{0}}}}{h_0-1}\Big)^{-2}.
\end{equation}
Thus phantom crossing is possible if either conditions (\ref{c1})
and (\ref{c2}) or (\ref{c3}) and (\ref{c4}) are satisfied. This
implies that the model of entropy corrected holographic dark energy
gives its equation of state across $-1$, consistent with the
Gauss-Bonnet model for the correspondence to be generically
applicable.

\section{Conclusions}
Within the different candidates to play the role of the dark energy,
the entropy-corrected holographic dark energy model, has emerged as
a possible model with EoS across $-1$ \cite{rev}. In the present
paper we have studied cosmological application of holographic dark
energy density in the Gauss-Bonnet gravity framework. By considering
the entropy-corrected holographic energy density as a dynamical
cosmological constant, we have obtained the equation of state for
the holographic energy density in the Gauss-Bonnet framework. After
that we have studied the conditions under which we can obtain a
correspondence between entropy-corrected holographic and
Gauss-Bonnet models of dark energy.
\subsubsection*{Acknowledgments} The work of M. R. Setare has been
supported by Research Institute for Astronomy and Astrophysics of
Maragha. We would like to thank the anonymous referee for useful
comments on this work.

\end{document}